\documentclass[]{PoS}
\newcommand{\squeezeup}{\vspace{-3.0mm}}
\newcommand{\GeV}{\ensuremath{\,\mathrm{GeV}}\xspace}

\usepackage{xspace}
\usepackage{graphicx}
\usepackage{color}
\usepackage{dcolumn}
\usepackage{bm}

\title{Beyond the $t$-channel Approximation: Next-to-Leading Order QCD Corrections to Electroweak Higgs Boson Plus Three Jet Production at the LHC}

\ShortTitle{NLO QCD Corrections to EW $Hjjj$ Production at the LHC}
        
\author{Francisco Campanario \\ 
             Theory Division, IFIC, University of Valencia-CSIC \\E-46100 Paterna, Valencia, Spain\\ 
             E-mail:  \email{francisco.campanario@ific.uv.es}}
             
\author{\speaker{Terrance M. Figy} \\ 
School of Physics and Astronomy, The University of Manchester \\Manchester, M13 9PL, United Kingdom \\
E-mail: \email{Terrance.Figy@hep.manchester.ac.uk}}

\author{Simon Pl\"atzer \\
Theory Group, DESY  \\D-22607 Hamburg, Germany\\
E-mail:\email{simon.plaetzer@desy.de}}

\author{Malin Sj\"odahl \\
Department of Astronomy and Theoretical Physics, Lund University \\ SE-22362 Lund, Sweden\\
\email{malin.sjodahl@thep.lu.se}}

\abstract{
In this talk we discuss the implementation of the full next-to-leading order QCD corrections to electroweak Higgs boson plus three jet production at the LHC within the {\tt Matchbox} framework of the {\tt Herwig++} event generator.  We also present numerical results for integrated cross sections and kinematic distributions.%}
\\
\hspace*{-1 cm}
\begin{picture}(0,0) \put(0,580){\large DESY 14-110, FTUV-14-0714, IFIC/14-46, LPN14-094, LU TP 14-25, MAN/HEP/2014/08}\end{picture}
}

\FullConference{Loops and Legs in Quantum Field Theory - LL 2014,\\
		 27 April - 2 May 2014 \\
		 Weimar, Germany }

\begin{document}

\section{Introduction}

The existence of a new boson with a mass in the range of $125$--$126$ GeV, 
with a spin most likely equal to zero and with even parity has been 
confirmed with increasing confidence in recent reports by the ATLAS and CMS Collaborations 
\cite{Aad:2012tfa,Chatrchyan:1471016, Aad:2013xqa,CMS-PAS-HIG-13-005}. 
Furthermore, the new particle exhibits production and decay rates 
similar to a Standard Model (SM) Higgs boson 
\cite{Higgs:1964pj,Higgs:1964ia,Englert:1964et,Guralnik:1964eu, Aad:2013wqa,CMS-PAS-HIG-13-016}.

Higgs boson production via vector boson fusion (VBF), i.e., 
the $t$-channel $\mathcal{O}(\alpha_{QED}^{3})$ merging of two weak bosons 
in the reaction $qq \to qq H$, is an essential channel at the LHC
for constraining Higgs boson couplings to gauge bosons and fermions.  
In the current experimental data from the LHC,  the ATLAS Collaboration finds 
$3 \sigma$  evidence \cite{Aad:2013wqa} for Higgs boson production via 
VBF and the CMS Collaboration finds $1.3\sigma$ evidence \cite{CMS-PAS-HIG-13-022}.
  
For this process the observation of two forward tagging jets 
is crucial for the reduction of background.   
Requiring, in addition, that there is no extra radiation within the rapidity 
gap between the forward tagging jets ~\cite{Barger:1994zq,Rainwater:1998kj,Rainwater:1999sd},  
i.e., imposing a central jet veto (CJV), suppresses standard QCD backgrounds, 
as well as Higgs production via gluon-gluon fusion in association with two jets 
(GF $Hjj$) ~\cite{Barger:1994zq,Rainwater:1998kj,Rainwater:1999sd}.

To exploit the CJV strategy for Higgs boson coupling measurements, 
it is therefore necessary to know the reduction due to the CJV accurately.  
Thus it is of interest to calculate the ratio of Higgs boson plus three 
jet (EW $Hjjj$) production (where the third jet is required to be between
the two tagging jets) to the inclusive Higgs boson plus two jet (EW $Hjj$)
cross section.

Recently the competing GF $Hjjj$ has been computed within the heavy top 
effective theory approximation to next-to-leading order (NLO) in perturbative 
QCD~\cite{Cullen:2013saa}. 
The heavy top effective theory approximation for $Hjj(j)$ has been validated 
against $Hjj(j)$ amplitudes where the top mass dependence has been kept 
in Refs.~\cite{Campanario:2013mga,DelDuca:2001eu}. 

Approximated results at NLO QCD for  EW VBF $Hjjj$ production were presented
sometime ago in~\cite{Figy:2007kv,Figy:2006vc}. There, the $t$-channel
approximation was used and additionally, the 
inclusion of pentagon and hexagon one-loop Feynman diagram topologies 
(Figure ~\ref{fig:ggf}, last two diagrams) and the corresponding 
real emission contributions were omitted and estimated to contribute at the
per-mille level.  Recently, parton-shower effects on EW VBF $Hjjj$ were investigated in Ref.~\cite{Jager:2014vna} within 
the $t$-channel approximation\footnote{Parton-shower effects were
  investigated also in Ref.~\cite{Alwall:2014hca}.}.
In view of the relevance to the determination of Higgs boson couplings, 
we will present results from \cite{PhysRevLett.111.211802,Campanario:2013nca}, where 
those approximations are lifted, and the full NLO QCD corrections to 
the $\mathcal{O}(\alpha_{s} \alpha_{EW}^{3}$) production of a Higgs boson 
in association of three jets is calculated for the first time. In this
proceedings, we show results for the inclusive sample and leave for future work
a thorough comparison with the VBF approximation.

The remainder of this proceedings is organized as follows: 
Details of the NLO calculation are presented in Section~\ref{details}. Numerical results and conclusions are shown in Section~\ref{results} and in
Section~\ref{concl}, respectively.

\squeezeup
\section{Calculational Details}
\label{details}

For the leading order (LO) $2 \to H+n$ ($n=2,3,4$) parton matrix elements, 
we employ the built--in spinor helicity library of the {\tt Matchbox} module 
in the {\tt Herwig++} event generator~\cite{Bahr:2008pv,Bellm:2013lba} to construct the 
full amplitude from hadronic currents~\cite{Platzer:2011bc}.  
The LO $2 \to H+n$ ($n=2,3,4$) parton matrix elements were also cross 
checked against {\tt Sherpa}~\cite{Gleisberg:2003xi,Gleisberg:2008ta}, 
{\tt VBFNLO}~\cite{Arnold:2011wj,Arnold:2012xn,Baglio:2014uba}, and 
{\tt Hawk}~\cite{Ciccolini:2007jr,Ciccolini:2007ec}. 
The Catani--Seymour dipole subtraction
terms \cite{Catani:1996vz} are generated automatically by the
{\tt Matchbox} module~\cite{Platzer:2011bc}, and for efficient generation 
of phase space points, we utilize a diagram-based multichannel phase 
space sampler~\cite{Platzer:2011bc}.

\begin{figure}[ht]
\begin{center}
\includegraphics[scale=0.65]{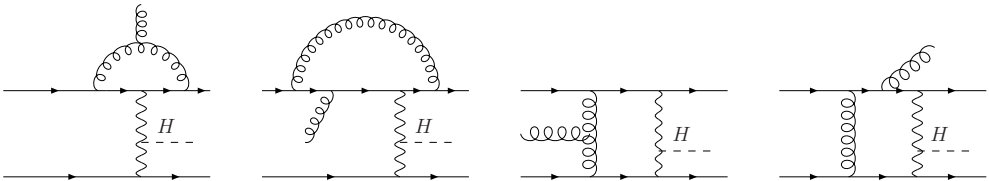}
\end{center}
\squeezeup
\caption{\label{fig:ggf}
A representative selection of one-loop Feynman diagram topologies for EW $Hjjj$ production.}  
\end{figure}

The computation of the interference of the virtual one--loop amplitude with 
Born amplitudes, is calculated with the aid of the helicity amplitude technique
described in Ref.~\cite{Hagiwara:1988pp}, using the program described in  Ref.~\cite{Campanario:2011cs}, which also provides an independent
version of the Born amplitudes, providing a valuable internal consistency
check of our implementation. A representative set of one--loop Feynman diagram topologies that contribute 
to the virtual corrections are depicted in Figure ~\ref{fig:ggf}. 
To evaluate the one--loop tensor coefficients, we use the
Passarino-Veltman approach \cite{Passarino:1978jh} up to four-point functions, and the Denner-Dittmaier 
scheme~\cite{Denner:2005nn}, following the layout and notation 
of~\cite{Campanario:2011cs}, to numerically evaluate the five and 
six point coefficients.
The one--loop scalar integrals are in turn evaluated using
the program {\tt OneLOop}~\cite{vanHameren:2010cp}. Complex masses and  
finite width effects in gauge boson propagators are calculated in the 
complex mass scheme ~\cite{Denner:2006ic,Nowakowski:1993iu}. 
The
resulting one--loop amplitudes for specific phase space points have been cross
checked against {\tt GoSam}~\cite{Cullen:2011ac}.

The numerical stability of our code, is tested by employing a Ward
identity check at each phase space point and each Feynman diagram 
\cite{Campanario:2011cs} -- at the cost of a small increase in computing
time. 
Upon failure of the Ward identity check, the amplitudes of the gauge 
related topology are set to zero.  The failure rate is at the 
per-mille level and hence under control. 
This method has also been successfully applied in other scattering processes 
with $2 \to 4$ kinematics~\cite{Campanario:2011ud,Campanario:2013qba}, however, in the work presented here, the method is applied to a process 
which involves loop propagators with complex masses for the first time.

The color structure associated with the computation of color correlated 
Born matrix elements has been performed by {\tt ColorFull}
\cite{Sjodahl:ColorFull} and cross checked against {\tt ColorMath}
\cite{Sjodahl:2012nk}. As a further check of our framework, we have implemented
the corresponding calculation of electroweak $Hjj$ production and, subsequently, 
performed cross checks against {\tt Hawk}~\cite{Ciccolini:2007jr,Ciccolini:2007ec} 
and VBFNLO~\cite{Arnold:2011wj,Arnold:2012xn,Baglio:2014uba}.

We refer to our implementation of the NLO corrections in perturbative QCD for 
electroweak Higgs boson plus two and three jet production in the 
{\tt Matchbox} framework as {\tt HJets++}.

\squeezeup
\section{Results}
\label{results}

In this section, we present results for a LHC of center-of-mass 
energy $\sqrt{s}=14$ TeV. 
%We use {\tt Herwig++}~\cite{Bahr:2008pv} to generate and analyze NLO events. 
Here, we do not include parton shower and harmonization effects in our 
simulations. Instead the matrix element partons are recombined into jets 
according to the anti-$k_{T}$ algorithm~\cite{Cacciari:2008gp} 
using {\tt FastJet} \cite{Cacciari:2011ma} with $D=0.4$ and $E$-scheme recombination.
We select events with at least three jets having transverse momentum $p_{T,j} \ge
20~\rm{GeV}$ and rapidity $|y_{j}| \le 4.5$ and order the jets according to their
transverse momentum.

We use the CT10~\cite{Lai:2010vv} 
parton distribution functions with $\alpha_s(M_Z)= 0.118$ at NLO, 
and the CTEQ6L1 set~\cite{Pumplin:2002vw} with $\alpha_s(M_Z)=0.130$ at LO. 
We use the five-flavor scheme for the running of $\alpha_s$.
We choose $m_Z=91.188 \GeV$, $m_W=80.419002 \GeV$,
$m_H=125 \GeV$ and $G_F=1.16637\times 10^{-5}\GeV^{-2}$ as electroweak
input parameters and derive the weak mixing angle $\sin \theta_{W}$
and $\alpha_{QED}$ from SM tree level relations. 
All
fermion masses (except the top quark) are set to zero and the CKM matrix is taken to be diagonal. 
Widths are fixed to the following values: $\Gamma_W=2.0476$ GeV and $\Gamma_Z=2.4414$ GeV.

\begin{figure}[]
\begin{center}
\includegraphics[scale=0.9]{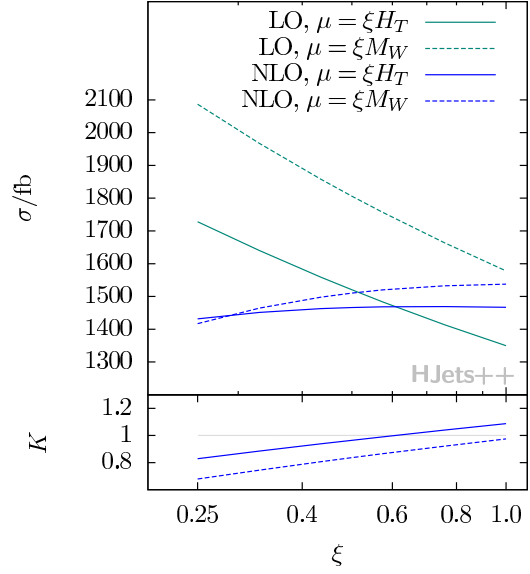}
\end{center}
\squeezeup
\caption{\label{fig:scale} The $Hjjj$ inclusive total cross section
  (in fb) at LO (cyan) and at NLO (blue) for the scale choices,
  $\mu=\xi M_{W}$ (dashed) and $\mu=\xi H_{T}$ (solid). 
  The lower panel displays the $K$-factor, $K=\sigma_{NLO}/\sigma_{LO}$ f
  or $\mu=\xi M_{W}$ (dashed) and $\mu=\xi H_{T}$ (solid).  }
\end{figure}

In Figure~\ref{fig:scale}, we show the LO and NLO total cross-sections
for inclusive cuts for different values of the factorization scale ($\mu_{F}$) and
renormalization scale ($\mu_{R}$), varied around the central scale, $\mu$ for two
different scale choices, $M_{W}/2$, and the scalar sum of the jet transverse
momenta, $H_{T}/2$ with $H_{T}=\sum_{j}
p_{T,j}$. In general, we find %a somewhat increased cross section and 
-- as expected -- a decreased scale dependence in the NLO results.  We also
note that the central values for the various scale choices are closer
to each other at NLO.  
The uncertainties, obtained by varying the
central value a factor two up and down, are around $25\%$ ($28\%$) at LO and
$2\%$ ($8\%$) at NLO using $H_{T}/2$ ($M_W/2$) as scale choice. 
For the scale choice $\mu = H_T/2$, we obtained
$\sigma_{LO}=1520(8)^{+208}_{-171}$ fb and 
$\sigma_{NLO}=1466(17)^{+1}_{-35}$ fb.  Studying differential
distributions, we find that these generally vary less using the scalar
transverse momentum sum choice, used from now on.

\begin{figure}[ht]
\begin{tabular}{cc}
\includegraphics[scale=0.62]{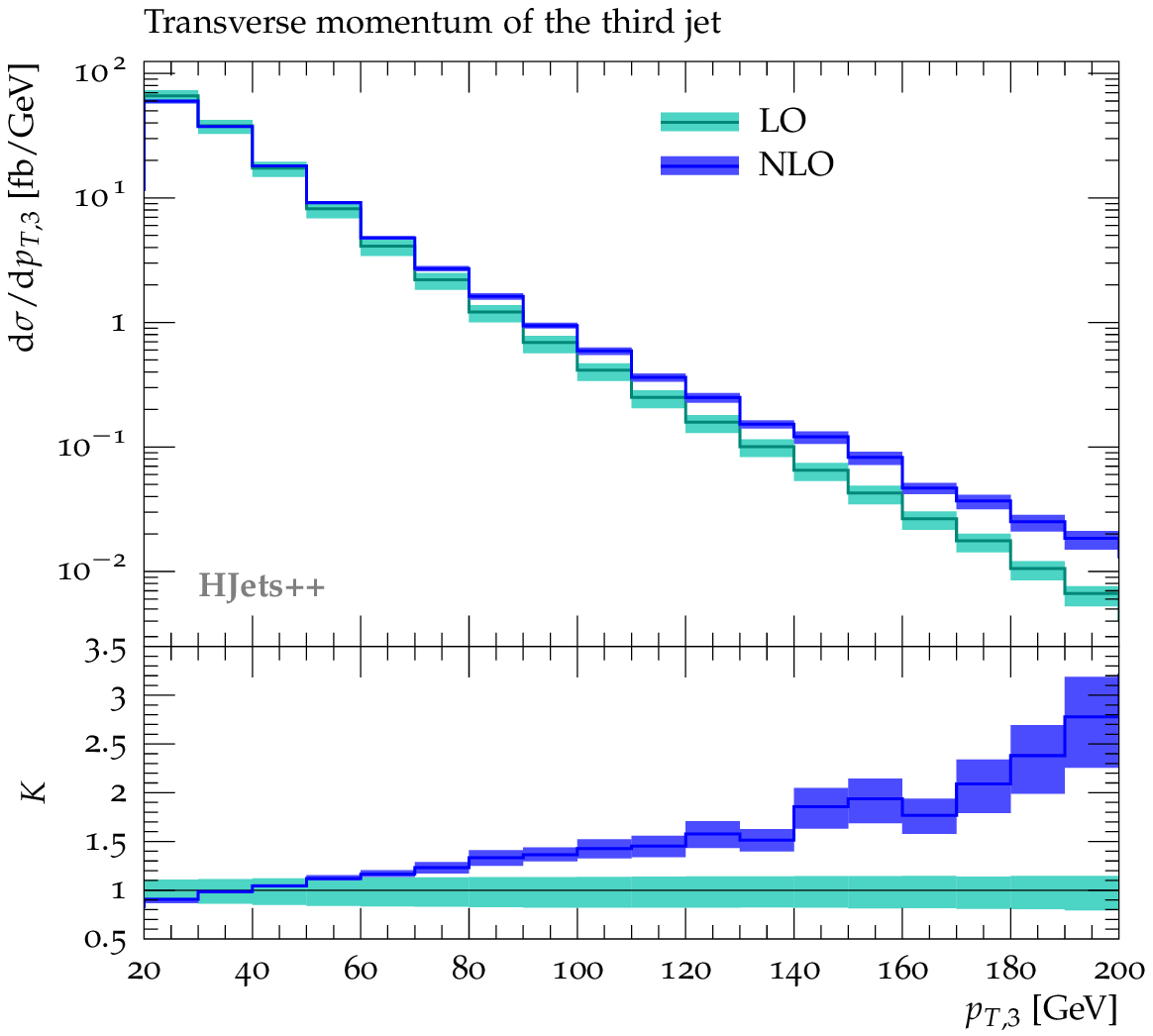} & \includegraphics[scale=0.62]{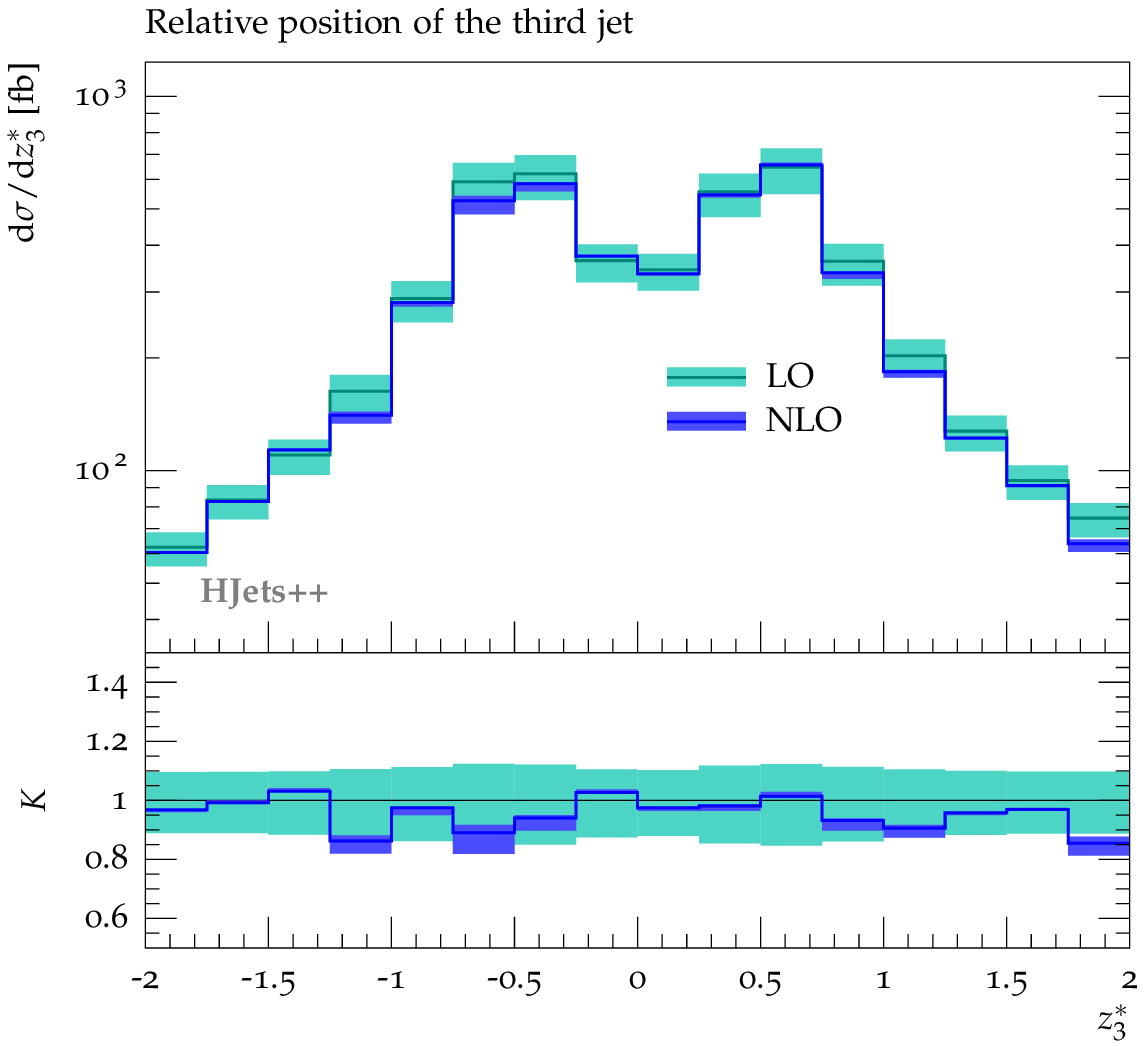} \\
\end{tabular}
\squeezeup
\caption{\label{fig:pt3} Differential cross section and $K$ factor for
  the $p_T$ of the third hardest jet (left) and the normalized centralized rapidity distribution of the third jet
  w.r.t. the tagging jets (right). Cuts are described
  in the text. The bands correspond to varying $\mu_F=\mu_R$ by
  factors 1/2 and 2 around the central value $H_{T}/2$.  }
\end{figure}

\begin{figure}[ht]
\begin{tabular}{cc}
\includegraphics[scale=0.62]{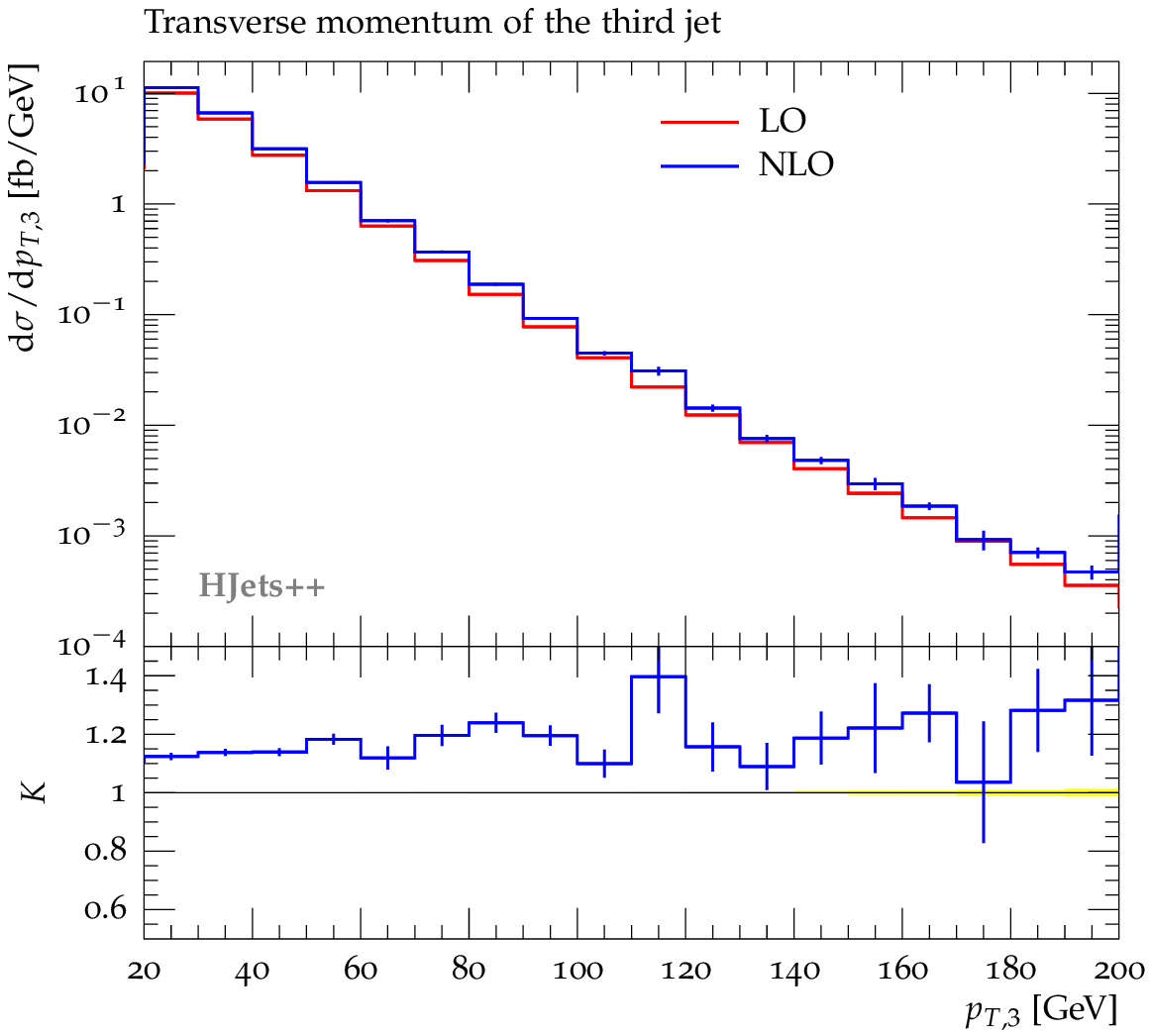} & \includegraphics[scale=0.62]{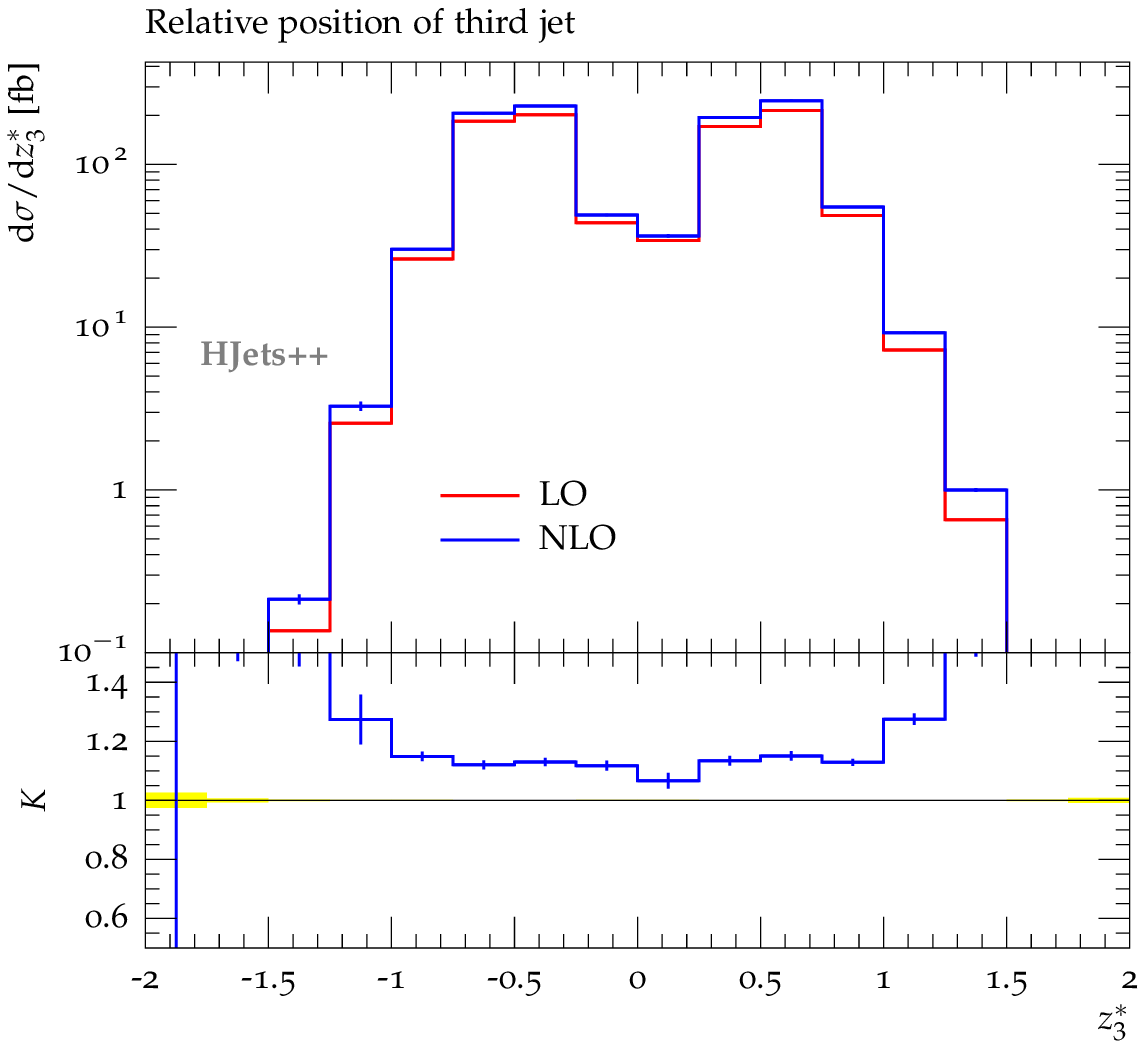}
\end{tabular}
\squeezeup
\caption{\label{fig:pt3vbf} Differential cross section and $K$ factor for
  the $p_T$ of the third hardest jet (left) and the normalized centralized 
  rapidity distribution of the third jet
  w.r.t. the tagging jets (right) with $\mu_{R}=\mu_{F}=H_{T}$. Beyond the inclusive cuts described in the text, we include the set of VBF cuts: $m_{12} = \sqrt{(p_{1}+p_{2})^2} > 600~{\rm GeV}$ and $|\Delta y_{12}| = |y_{1} - y_{2}|> 4.0$.}
\end{figure}

On the left-hand side of Figure~\ref{fig:pt3}, the differential 
distribution of the third jet, (i.e., the jet which would 
be vetoed in a CJV analysis), is shown. Here we find
large $K$ factors in the high energy tail of the transverse momentum
distribution. 
However, when VBF cuts \footnote{For the VBF cuts we have chosen to include the following cuts 
in addition to the inclusive cuts described in the main text : 
$m_{12} = \sqrt{(p_{1}+p_{2})^2} > 600~{\rm GeV}$ and $|\Delta y_{12}| =
|y_{1} - y_{2}|> 4.0$}  
are imposed 
the $K$ factor is almost flat as a function of the transverse momentum of the 
third jet (see the left-hand side of Figure~\ref{fig:pt3vbf}).
On the right-hand side of Figure~\ref{fig:pt3}, we show 
the normalized centralized rapidity distribution of the third jet 
w.r.t. the tagging jets, $z^{*}_{3}=(y_3-\frac{1}{2}(y_1+y_2))/(y_1-y_2)$. 
This variable, showing how the third jet tends to accompany one 
of the leading jets appearing at $1/2$ and $-1/2$ respectively, 
beautifully displays the VBF nature present in the process.

This effect is even more pronounced when VBF cuts are applied 
(see Figure \ref{fig:pt3vbf}), and should be contrasted with the gluon 
fusion production mechanism where
QCD radiation in the rapidity gap region between the leading two jets is much
more common~\cite{Forshaw:2007vb,Cox:2010ug,Campanario:2013mga,Cullen:2013saa}.

\squeezeup
\section{Conclusions}
\label{concl}
\squeezeup
In this proceedings, we have presented complete results at NLO QCD for 
electroweak Higgs boson production in association with three jets.
We have found that the NLO corrections to the total inclusive cross 
section are moderate for inclusive cuts using the scale choice of $H_{T}/2$. 
However, for the scale choice of $M_{W}/2$, the NLO corrections can be more significant.
The scale uncertainty decreases from around $25\%$($28\%$) at LO down 
to about $2\%(${$8\%$) at NLO using the scale choice of $H_{T}/2$ ($M_{W}/2$). 
We have also presented numerical results showing the impact of 
VBF selection cuts on the transverse momentum of the third jet, 
$p_{T,3}$, and its relative position w.r.t. the two leading jets, 
$z_{3}^{*}$.% at NLO in perturbative QCD.

\squeezeup
\acknowledgments{We are grateful to Ken Arnold for contributions at an early stage of this project and to Mike Seymour and Jeff Forshaw for valuable discussions on the subject. F.C. is funded by a Marie Curie fellowship (PIEF-GA-2011- 298960) and partially by MINECO (FPA2011-23596) and by LHCPhenonet (PITN-GA-2010-264564). T.F. would like to thank the North American Foundation for The University of Manchester and George Rigg for their financial support. S.P. has been supported in part by the Helmholtz Alliance "Physics at the Terascale" and M.S. was supported by the Swedish Research Council, contract number 621-2010-3326. Numerical computations presented in this proceedings were performed via PhenoGrid using GridPP infrastructure and via the DESY Computing Cluster called Bird.}
\squeezeup
\bibliographystyle{JHEP}
%\bibliography{nlo3jets-PoS}
\bibliography{mybib}

\providecommand{\href}[2]{#2}\begingroup\raggedright\begin{thebibliography}{10}

\bibitem{Aad:2012tfa}
{\bf ATLAS Collaboration} Collaboration, G.~Aad et~al., {\it {Observation of a
  new particle in the search for the Standard Model Higgs boson with the ATLAS
  detector at the LHC}},  {\em Phys.~Lett.} {\bf B716} (2012) 1--29,
  [\href{http://xxx.lanl.gov/abs/1207.7214}{{\tt arXiv:1207.7214}}].

\bibitem{Chatrchyan:1471016}
S.~Chatrchyan et~al., {\it {Observation of a new boson at a mass of $125 {\rm
  GeV}$ with the CMS experiment at the LHC}},  {\em Phys.~Lett.~B} {\bf 716}
  (Jul, 2012) 30--61. 59 p.

\bibitem{Aad:2013xqa}
{\bf ATLAS Collaboration} Collaboration, G.~Aad et~al., {\it {Evidence for the
  spin-0 nature of the Higgs boson using ATLAS data}},
  \href{http://xxx.lanl.gov/abs/1307.1432}{{\tt arXiv:1307.1432}}.

\bibitem{CMS-PAS-HIG-13-005}
{\it {Combination of standard model Higgs boson searches and measurements of
  the properties of the new boson with a mass near $125~{\rm GeV}$}},  Tech.
  Rep. CMS-PAS-HIG-13-005, CERN, Geneva, 2013.

\bibitem{Higgs:1964pj}
P.~W. Higgs, {\it {Broken Symmetries and the Masses of Gauge Bosons}},  {\em
  Phys.~Rev.~Lett.} {\bf 13} (1964) 508--509.

\bibitem{Higgs:1964ia}
P.~W. Higgs, {\it {Broken symmetries, massless particles and gauge fields}},
  {\em Phys.~Lett.} {\bf 12} (1964) 132--133.

\bibitem{Englert:1964et}
F.~Englert and R.~Brout, {\it {Broken Symmetry and the Mass of Gauge Vector
  Mesons}},  {\em Phys.~Rev.~Lett.} {\bf 13} (1964) 321--323.

\bibitem{Guralnik:1964eu}
G.~Guralnik, C.~Hagen, and T.~Kibble, {\it {Global Conservation Laws and
  Massless Particles}},  {\em Phys.~Rev.~Lett.} {\bf 13} (1964) 585--587.

\bibitem{Aad:2013wqa}
{\bf ATLAS Collaboration} Collaboration, G.~Aad et~al., {\it {Measurements of
  Higgs boson production and couplings in diboson final states with the ATLAS
  detector at the LHC}},  \href{http://xxx.lanl.gov/abs/1307.1427}{{\tt
  arXiv:1307.1427}}.

\bibitem{CMS-PAS-HIG-13-016}
{\it Properties of the observed higgs-like resonance using the diphoton
  channel},  Tech. Rep. CMS-PAS-HIG-13-016, CERN, Geneva, 2013.

\bibitem{CMS-PAS-HIG-13-022}
{\it {Update of the search for the Standard Model Higgs boson decaying into
  $WW$ in the vector boson fusion production channel}},  Tech. Rep.
  CMS-PAS-HIG-13-022, CERN, Geneva, 2013.

\bibitem{Barger:1994zq}
V.~D. Barger, R.~Phillips, and D.~Zeppenfeld, {\it {Mini - jet veto: A Tool for
  the heavy Higgs search at the LHC}},  {\em Phys.~Lett.} {\bf B346} (1995)
  106--114, [\href{http://xxx.lanl.gov/abs/hep-ph/9412276}{{\tt
  hep-ph/9412276}}].

\bibitem{Rainwater:1998kj}
D.~L. Rainwater, D.~Zeppenfeld, and K.~Hagiwara, {\it {Searching for
  $H\to\tau^+\tau^-$ in weak boson fusion at the CERN LHC}},  {\em Phys.~Rev.}
  {\bf D59} (1998) 014037, [\href{http://xxx.lanl.gov/abs/hep-ph/9808468}{{\tt
  hep-ph/9808468}}].

\bibitem{Rainwater:1999sd}
D.~L. Rainwater and D.~Zeppenfeld, {\it {Observing $H\to W^*W^* \to e^\pm
  \mu\mp \not{p}_T$ in weak boson fusion with dual forward jet tagging at the
  CERN LHC}},  {\em Phys.~Rev.} {\bf D60} (1999) 113004,
  [\href{http://xxx.lanl.gov/abs/hep-ph/9906218}{{\tt hep-ph/9906218}}].

\bibitem{Cullen:2013saa}
G.~Cullen, H.~van Deurzen, N.~Greiner, G.~Luisoni, P.~Mastrolia, et~al., {\it
  {NLO QCD corrections to Higgs boson production plus three jets in gluon
  fusion}},  {\em Phys.~Rev.~Lett.} {\bf 111} (2013) 131801,
  [\href{http://xxx.lanl.gov/abs/1307.4737}{{\tt arXiv:1307.4737}}].

\bibitem{Campanario:2013mga}
F.~Campanario and M.~Kubocz, {\it {Higgs boson production in association with
  three jets via gluon fusion at the LHC: Gluonic contributions}},  {\em
  Phys.~Rev.} {\bf D88} (2013) 054021,
  [\href{http://xxx.lanl.gov/abs/1306.1830}{{\tt arXiv:1306.1830}}].

\bibitem{DelDuca:2001eu}
V.~Del~Duca, W.~Kilgore, C.~Oleari, C.~Schmidt, and D.~Zeppenfeld, {\it Higgs +
  2 jets via gluon fusion},  {\em Phys.~Rev.~Lett.} {\bf 87} (2001) 122001,
  [\href{http://xxx.lanl.gov/abs/hep-ph/0105129}{{\tt hep-ph/0105129}}].

\bibitem{Figy:2007kv}
T.~Figy, V.~Hankele, and D.~Zeppenfeld, {\it {Next-to-leading order QCD
  corrections to Higgs plus three jet production in vector-boson fusion}},
  {\em JHEP} {\bf 0802} (2008) 076,
  [\href{http://xxx.lanl.gov/abs/0710.5621}{{\tt arXiv:0710.5621}}].

\bibitem{Figy:2006vc}
T.~M. Figy, {\it {NLO QCD corrections to the jet activity in Higgs boson
  production via vector-boson fusion}},  {\em Ph.D. Thesis} (2006).

\bibitem{Jager:2014vna}
B.~Jager, F.~Schissler, and D.~Zeppenfeld, {\it {Parton-shower effects on Higgs
  boson production via vector-boson fusion in association with three jets}},
  \href{http://xxx.lanl.gov/abs/1405.6950}{{\tt arXiv:1405.6950}}.

\bibitem{Alwall:2014hca}
J.~Alwall, R.~Frederix, S.~Frixione, V.~Hirschi, F.~Maltoni, et~al., {\it {The
  automated computation of tree-level and next-to-leading order differential
  cross sections, and their matching to parton shower simulations}},
  \href{http://xxx.lanl.gov/abs/1405.0301}{{\tt arXiv:1405.0301}}.

\bibitem{PhysRevLett.111.211802}
F.~Campanario, T.~M. Figy, S.~Pl\"atzer, and M.~Sj\"odahl, {\it Electroweak
  higgs boson plus three jet production at next-to-leading-order qcd},  {\em
  Phys.~Rev.~Lett.} {\bf 111} (Nov, 2013) 211802.

\bibitem{Campanario:2013nca}
F.~Campanario, T.~M. Figy, S.~Platzer, and M.~Sjodahl, {\it {NLO QCD
  Corrections to Electroweak Higgs Boson Plus Three Jet Production at the
  LHC}},  {\em PoS} {\bf RADCOR2013} (2014) 042,
  [\href{http://xxx.lanl.gov/abs/1311.5455}{{\tt arXiv:1311.5455}}].

\bibitem{Bahr:2008pv}
{M. B\"ahr et al.}, {\it {Herwig++ Physics and Manual}},  {\em Eur.~Phys.~J.}
  {\bf C58} (2008) 639--707, [\href{http://xxx.lanl.gov/abs/0803.0883}{{\tt
  arXiv:0803.0883}}].

\bibitem{Bellm:2013lba}
J.~Bellm, S.~Gieseke, D.~Grellscheid, A.~Papaefstathiou, S.~Platzer, et~al.,
  {\it {Herwig++ 2.7 Release Note}},
  \href{http://xxx.lanl.gov/abs/1310.6877}{{\tt arXiv:1310.6877}}.

\bibitem{Platzer:2011bc}
S.~Platzer and S.~Gieseke, {\it {Dipole Showers and Automated NLO Matching in
  Herwig++}},  {\em Eur.~Phys.~J.} {\bf C72} (2012) 2187,
  [\href{http://xxx.lanl.gov/abs/1109.6256}{{\tt arXiv:1109.6256}}].

\bibitem{Gleisberg:2003xi}
T.~Gleisberg et~al., {\it {SHERPA 1.alpha, a proof-of-concept version}},  {\em
  JHEP} {\bf 02} (2004) 056,
  [\href{http://xxx.lanl.gov/abs/hep-ph/0311263}{{\tt hep-ph/0311263}}].

\bibitem{Gleisberg:2008ta}
T.~Gleisberg et~al., {\it {Event generation with SHERPA 1.1}},  {\em JHEP} {\bf
  02} (2009) 007, [\href{http://xxx.lanl.gov/abs/0811.4622}{{\tt
  arXiv:0811.4622}}].

\bibitem{Arnold:2011wj}
K.~Arnold, J.~Bellm, G.~Bozzi, M.~Brieg, F.~Campanario, et~al., {\it {VBFNLO: A
  Parton Level Monte Carlo for Processes with Electroweak Bosons -- Manual for
  Version 2.5.0}},  \href{http://xxx.lanl.gov/abs/1107.4038}{{\tt
  arXiv:1107.4038}}.

\bibitem{Arnold:2012xn}
K.~Arnold, J.~Bellm, G.~Bozzi, F.~Campanario, C.~Englert, et~al., {\it {Release
  Note -- Vbfnlo-2.6.0}},  \href{http://xxx.lanl.gov/abs/1207.4975}{{\tt
  arXiv:1207.4975}}.

\bibitem{Baglio:2014uba}
J.~Baglio, J.~Bellm, F.~Campanario, B.~Feigl, J.~Frank, et~al., {\it {Release
  Note - VBFNLO 2.7.0}},  \href{http://xxx.lanl.gov/abs/1404.3940}{{\tt
  arXiv:1404.3940}}.

\bibitem{Ciccolini:2007jr}
M.~Ciccolini, A.~Denner, and S.~Dittmaier, {\it {Strong and electroweak
  corrections to the production of Higgs + 2jets via weak interactions at the
  LHC}},  {\em Phys.~Rev.~Lett.} {\bf 99} (2007) 161803,
  [\href{http://xxx.lanl.gov/abs/0707.0381}{{\tt arXiv:0707.0381}}].

\bibitem{Ciccolini:2007ec}
M.~Ciccolini, A.~Denner, and S.~Dittmaier, {\it {Electroweak and QCD
  corrections to Higgs production via vector-boson fusion at the LHC}},  {\em
  Phys.~Rev.} {\bf D77} (2008) 013002,
  [\href{http://xxx.lanl.gov/abs/0710.4749}{{\tt arXiv:0710.4749}}].

\bibitem{Catani:1996vz}
{S. Catani and M.H. Seymour}, {\it {A general algorithm for calculating jet
  cross sections in NLO QCD}},  {\em Nucl.~Phys.} {\bf B485} (1997) 291--419,
  [\href{http://xxx.lanl.gov/abs/hep-ph/9605323}{{\tt hep-ph/9605323}}].

\bibitem{Hagiwara:1988pp}
K.~Hagiwara and D.~Zeppenfeld, {\it {Amplitudes for Multiparton Processes
  Involving a Current at e+ e-, e+- p, and Hadron Colliders}},  {\em
  Nucl.~Phys.} {\bf B313} (1989) 560.

\bibitem{Campanario:2011cs}
F.~Campanario, {\it {Towards $pp \to VVjj$ at NLO QCD: Bosonic contributions to
  triple vector boson production plus jet}},  {\em JHEP} {\bf 1110} (2011) 070,
  [\href{http://xxx.lanl.gov/abs/1105.0920}{{\tt arXiv:1105.0920}}].

\bibitem{Passarino:1978jh}
G.~Passarino and M.~Veltman, {\it {One Loop Corrections for e+ e- Annihilation
  Into mu+ mu- in the Weinberg Model}},  {\em Nucl.~Phys.} {\bf B160} (1979)
  151.

\bibitem{Denner:2005nn}
A.~Denner and S.~Dittmaier, {\it {Reduction schemes for one-loop tensor
  integrals}},  {\em Nucl.~Phys.} {\bf B734} (2006) 62--115,
  [\href{http://xxx.lanl.gov/abs/hep-ph/0509141}{{\tt hep-ph/0509141}}].

\bibitem{vanHameren:2010cp}
A.~van Hameren, {\it {OneLOop: For the evaluation of one-loop scalar
  functions}},  {\em Comput.~Phys.~Commun.} {\bf 182} (2011) 2427--2438,
  [\href{http://xxx.lanl.gov/abs/1007.4716}{{\tt arXiv:1007.4716}}].

\bibitem{Denner:2006ic}
A.~Denner and S.~Dittmaier, {\it {The Complex-mass scheme for perturbative
  calculations with unstable particles}},  {\em Nucl.~Phys.~Proc.~Suppl.} {\bf
  160} (2006) 22--26, [\href{http://xxx.lanl.gov/abs/hep-ph/0605312}{{\tt
  hep-ph/0605312}}].

\bibitem{Nowakowski:1993iu}
M.~Nowakowski and A.~Pilaftsis, {\it {On gauge invariance of Breit-Wigner
  propagators}},  {\em Z.~Phys.} {\bf C60} (1993) 121--126,
  [\href{http://xxx.lanl.gov/abs/hep-ph/9305321}{{\tt hep-ph/9305321}}].

\bibitem{Cullen:2011ac}
G.~Cullen, N.~Greiner, G.~Heinrich, G.~Luisoni, P.~Mastrolia, et~al., {\it
  {Automated One-Loop Calculations with GoSam}},  {\em Eur.~Phys.~J.} {\bf C72}
  (2012) 1889, [\href{http://xxx.lanl.gov/abs/1111.2034}{{\tt
  arXiv:1111.2034}}].

\bibitem{Campanario:2011ud}
F.~Campanario, C.~Englert, M.~Rauch, and D.~Zeppenfeld, {\it {Precise
  predictions for W $\gamma \gamma +$jet production at hadron colliders}},
  {\em Phys.~Lett.} {\bf B704} (2011) 515--519.

\bibitem{Campanario:2013qba}
F.~Campanario, M.~Kerner, L.~D. Ninh, and D.~Zeppenfeld, {\it {WZ production in
  association with two jets at NLO in QCD}},  {\em Phys.~Rev.~Lett. 111,} {\bf
  052003} (2013) [\href{http://xxx.lanl.gov/abs/1305.1623}{{\tt
  arXiv:1305.1623}}].

\bibitem{Sjodahl:ColorFull}
M.~Sjodahl, {\it {ColorFull -- A C++ package for color space calculations}},
  {\em http://colorfull.hepforge.org/}.

\bibitem{Sjodahl:2012nk}
M.~Sjodahl, {\it {ColorMath - A package for color summed calculations in
  SU(Nc)}},  {\em Eur.~Phys.~J.} {\bf C73} (2013) 2310,
  [\href{http://xxx.lanl.gov/abs/1211.2099}{{\tt arXiv:1211.2099}}].

\bibitem{Cacciari:2008gp}
M.~Cacciari, G.~P. Salam, and G.~Soyez, {\it {The Anti-k(t) jet clustering
  algorithm}},  {\em JHEP} {\bf 0804} (2008) 063,
  [\href{http://xxx.lanl.gov/abs/0802.1189}{{\tt arXiv:0802.1189}}].

\bibitem{Cacciari:2011ma}
M.~Cacciari, G.~P. Salam, and G.~Soyez, {\it {FastJet User Manual}},  {\em
  Eur.~Phys.~J.} {\bf C72} (2012) 1896,
  [\href{http://xxx.lanl.gov/abs/1111.6097}{{\tt arXiv:1111.6097}}].

\bibitem{Lai:2010vv}
H.-L. Lai, M.~Guzzi, J.~Huston, Z.~Li, P.~M. Nadolsky, et~al., {\it {New parton
  distributions for collider physics}},  {\em Phys.~Rev.} {\bf D82} (2010)
  074024, [\href{http://xxx.lanl.gov/abs/1007.2241}{{\tt arXiv:1007.2241}}].

\bibitem{Pumplin:2002vw}
J.~Pumplin, D.~Stump, J.~Huston, H.~Lai, P.~M. Nadolsky, et~al., {\it {New
  generation of parton distributions with uncertainties from global QCD
  analysis}},  {\em JHEP} {\bf 0207} (2002) 012,
  [\href{http://xxx.lanl.gov/abs/hep-ph/0201195}{{\tt hep-ph/0201195}}].

\bibitem{Forshaw:2007vb}
J.~R. Forshaw and M.~Sjodahl, {\it {Soft gluons in Higgs plus two jet
  production}},  {\em JHEP} {\bf 0709} (2007) 119,
  [\href{http://xxx.lanl.gov/abs/0705.1504}{{\tt arXiv:0705.1504}}].

\bibitem{Cox:2010ug}
B.~E. Cox, J.~R. Forshaw, and A.~D. Pilkington, {\it {Extracting Higgs boson
  couplings using a jet veto}},  {\em Phys.~Lett.} {\bf B696} (2011) 87--91,
  [\href{http://xxx.lanl.gov/abs/1006.0986}{{\tt arXiv:1006.0986}}].

\end{thebibliography}\endgroup

\end{document}